# Interface Characteristics at an Organic/Metal Junction: Pentacene on Cu Stepped Surfaces

Jeronimo Matos and Abdelkader Kara*
Department of Physics, University of Central Florida, Orlando, Florida 32816, USA

## Abstract

The adsorption of pentacene on Cu(221), Cu(511) and Cu(911) is investigated using density functional theory (DFT) with the self-consistent inclusion of van der Waals (vdW) interactions. Cu(211) is a vicinal of Cu(111) while Cu(511) and (911) are vicinals of Cu(100). For all the three surfaces, we found pentacene to prefer to adsorb parallel to the surface and near the steps. The addition of vdW interactions resulted in an enhancement in adsorption energies, with reference to the PBE functional, of around 2 eV. With vdWs inclusion, the adsorption energies were found to be 2.98 eV, 3.20 eV and 3.49 eV for Cu(211), Cu(511) and Cu(911) respectively. These values reflect that pentacene adsorbs stronger on (100) terraces with a preference for larger terraces. The molecule tilts upon adsorption with a small tilt angle on the (100) vicinals (about a few degrees) as compared to a large one on Cu(221) where the tilt angle is found to be about $20^{o}$. We find that the adsorption results in a net charge transfer to the molecule of ~1 electron, for all surfaces.

Keywords: vicinal surfaces; pentacene; density functional theory; van der Waals

# I. Introduction

Organic molecular films are widely used in the production of microelectronic devices. The pursuit of future technologies, with improved capabilities, performance, and efficiency, influences the study of organic films beginning at the nanoscale. One such example, the self-assembly of nanostructures, has been identified in recent years as a valuable aspect of fabrication methods for future electronic devices. To support these efforts with the "bottom up" approach, theoretical methods using density functional theory (DFT) with van der Waals (vdW) interactions are being employed to study these materials at the interface.[1-7]

In particular, the adsorption of pentacene on metal surfaces has been the subject of several theoretical and experimental studies, owed in part to the value presented by pentacene and pentacene-based molecules for low-cost organic photovoltaic applications.[8-10] The purpose of this study is two-fold: to study the adsorption of pentacene on Cu(511), Cu(223), and Cu(911) and to assess the influence of vdW interactions on the description of this adsorption using density functional theory.

At the organic semiconductor-metal interface, the surface geometry characteristics of the substrate, such has symmetry and distances between features on the metal surface are reflected in the ordering of pentacene films. The STM study by Lagoute, et al., reported that pentacene molecules, adsorbed on Cu(111) at low temperature and low coverage, align with the close packed rows of substrate atoms and adsorb with the molecule's center over the HCP cite.[11] STM and Low Energy Electron Diffraction (LEED) studies of pentacene on Cu(110) report a coverage-dependent multiphase arrangement of the molecules, which above 0.8 ML was observed to form long range ordered structures with the molecular axis aligned in the [1-10] substrate direction.[12-14]

Vicinal surfaces possess intrinsic strain near their step edges where low coordinated atoms are characterized by shorter bonds when compared to higher coordinated atoms in the bulk material or even at the terrace.[15-19] The preferential adsorption of molecules on step edges has been reported for a variety of systems.[20-22] This phenomenon facilitates the formation of ordered structures.[23] Furthermore, on vicinal surfaces of Copper, complex ordered structures pentacene, different from those found on low miller index surfaces, have been studied. The STM study by Fanetti *et al*., measured 25 ML films of pentacene on Cu(911) and found rippled morphologies of pentacene, with alternating mosaic and stair-like regions.[24] At a 1 ML coverage the pentacene molecules form long-range ordered chains along the step edge of Cu(911).[23] Contrastingly, on Cu(711), an absence of template induced ordering of multilayer films of pentacene was reported in an STM study by Gotzen *et al.*; although ordered chains of molecules were observed at 1 ML coverage, additional layers resulted in the growth of pyramidal islands.[25] For these reasons, we are motivated to study the effect of vdW interactions on the adsorption of pentacene on copper vicinal surfaces of the (100) and (111) type, with varying terrace widths.

For the adsorption of pentacene on Cu(221), Cu(511), and Cu(911) calculated using density functional theory with the inclusion of vdW interactions, we investigate the strength of the interaction between the molecule and the support as characterized by the criteria established by Yildirim, *et al.*[7] for chemisorption. Namely, strong chemisorption is characterized by strong binding to the support, changes in the geometry of the adsorbate and adsorbent, changes in the electronic structure of the surface, net charge transfer and the formation of an interface state. By presenting these results for pentacene on Cu(221), Cu(511), and Cu(911), with and without the inclusion of vdW interactions, we establish a basis for analyzing the effect of terrace width and vdW interaction on the description of the adsorption. In the following sections we will discuss the details of our calculations, present the adsorption energy and adsorption geometry results as well as the changes to the electronic structure of the substrate upon adsorption of the molecule.

## II. Computational Details

We perform density functional theory calculations on the adsorption of pentacene on Cu(511), Cu(911) and Cu(221) using the projector augmented wave method (PAW)[26, 27] implemented with the Vienna Ab-initio Simulation Package (VASP).[28, 29] Density functional theory calculations, conducted with the widely used local density approximation (LDA) and generalized gradient approximation (GGA), are unable to properly account for long range non-local interactions. Recently, several methods have been developed to include vdW interactions into DFT. For one method developed by Dion *et al.*[30-33] (vdW-DF), vdW interactions are included self consistently as an exchange-correlation functional. In this study, the exchange and correlation energy is calculated using the functionals introduced by Perdew *et al.*,[34] (PBE) and Klimes *et al.*[33] (optB88-vdW) to assess the effect of vdW interaction on the description of the adsorption. This exchange-correlation functional, from the self-consistent family of vdW-DF functionals, has been shown to be in good agreement with some of the available experimental results for the adsorption of benzene on metal surfaces,[1, 4, 6] and sexithiophene on Ag(110).[5]

The calculations are performed with a 400 eV plane wave energy cutoff. We use a 0.03 eV/Å force criterion for the convergence of our calculations. The Brillouin zone was sampled using a 5x3x1 k-mesh, for the Cu(511) surface, and 6x3x1 k-mesh, for the Cu(221) and Cu(911) surfaces.

The dimensions of the supercells are one step-terrace by seven copper atoms and at least 20 Å of vacuum vertically between copper slabs. Note that for comparison, the Cu(511) and Cu(911) surfaces are vicinal surfaces with (100) terraces while the Cu(221) surface has a (111) surface as its terrace (see Figure 1). The molecule in the gas phase and the clean substrate are allowed to relax completely before bringing the two together. We use the calculated bulk lattice constant 3.635 Å (PBE) and 3.626 Å (optB88-vdW) that we have reported in previous studies.[6] For the Pc/Cu system, the bottom layers of the copper slab are fixed at their bulk truncated positions. The pentacene molecule is initially placed near the step, at a height of 3 Å, with the molecular plane parallel to the copper terrace plane. Pentacene is only adsorbed on one side of the slab. We calculate the adsorption energy ($E_{ads}$) of pentacene on the copper surfaces as follows:

$$E_{ads} = -(E_{mol/sub} - E_{mol} - E_{sub})$$

were $E_{mol}$ is the total energy of the molecule in the gas phase, $E_{sub}$ is the total energy of the clean substrate, and $E_{mol/sub}$ is the total energy of the molecule-substrate system.

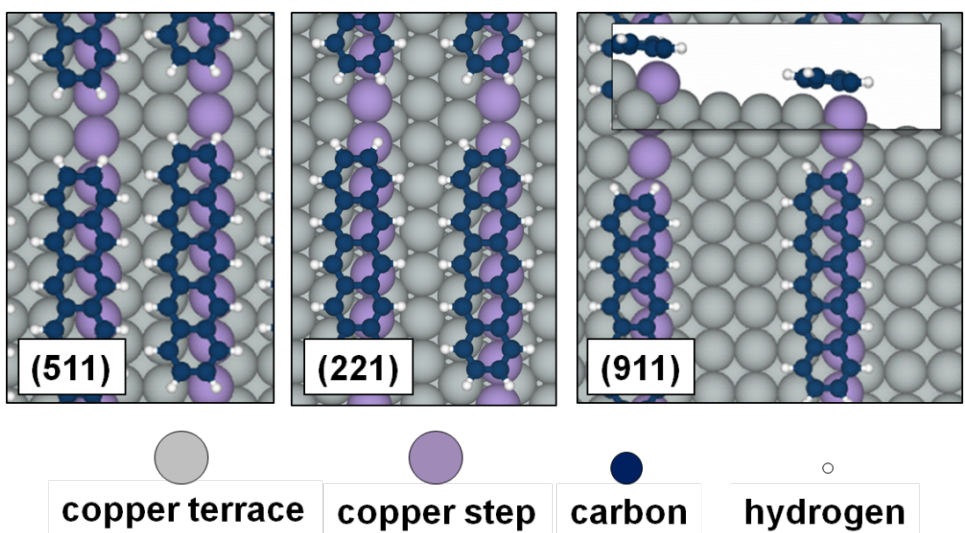

**Figure 1.** Adsorption of pentacene on Cu(511) [left], Cu(221) [center], and Cu(911) [right]. The copper atoms that make up the step edge are shown in purple as a guide for the eye.

## III. Results and Discussion

**III.1. Adsorption geometry and adsorption energies.**

The adsorption energies of pentacene on Cu(511), Cu(911), and Cu(221), using PBE and optB88-vdW, are reported in Table 1. We find that the adsorption energy is enhanced by about 2 eV with the addition of vdW interactions when compared to the results obtained using the PBE functional. The greatest enhancement in the adsorption energy corresponds to the adsorption of pentacene on Cu(511), while the least was that on Cu(221). The enhancement in adsorption energy provided by the optB88-vdW functional has been previously reported by our group for the adsorption of benzene, sexithiophene, and olympicene on transition metal surfaces with low miller indexes.[1, 5-7] We find that the trend in adsorption energy between the different stepped surfaces is different using PBE [Pc/Cu(511) < Pc/Cu(221) < Pc/Cu(911)] and optB88-vdW [Pc/Cu(221) < Pc/Cu(511) < Pc/Cu(911)].

**Table 1.** Adsorption Energy ($E_{ads}$), average adsorption height of carbon atoms in pentacene ($H_{ads}$), bending of the molecule ($\Delta z$), tilting of the molecule ($\alpha$), and buckling of the first layer surface copper atoms.

| Method | Surface | $E_{ads}$ (eV) | $H_{ads}$ (Å) | $\Delta z$ (Å) | $\alpha$ (°) | Buckling (Å) |
|---|---|---|---|---|---|---|
| PBE | (511) | 0.99 | 2.40 | 0.68 | -3.9 | 0.16 |
|  | (911) | 1.49 | 2.21 | 0.24 | -4.3 | 0.18 |
|  | (221) | 1.05 | 2.43 | 0.28 | -20.4 | 0.16 |
| optB88-vdW | (511) | 3.20 | 2.36 | 0.49 | -3.8 | 0.15 |
|  | (911) | 3.49 | 2.24 | 0.24 | -4.1 | 0.17 |
|  | (221) | 2.98 | 2.37 | 0.41 | -9.3 | 0.14 |

The adsorption heights, bending ($\Delta z$) and tilting ($\alpha$) of the pentacene molecule, and buckling of the first substrate layer upon adsorption on the Cu(511), Cu(911) and Cu(221) surfaces can also be found in Table 1. The adsorption height is calculated as the average distance between a carbon atom in the molecule and the plane containing the first layer copper atoms. We report a decrease in adsorption height with the use of optB88-vdW compared to PBE of 0.04 and 0.06 Å, for the adsorption of pentacene on Cu(511) and Cu(221), respectively. For the case of pentacene on Cu(911) we find a slight increase in the average molecule carbon height. It is interesting to note that even though the average height of carbon atoms in pentacene decreases, when using optB88-vdW, the smallest height between a carbon atom in the molecule and the surface plane increases by about 0.1 Å.

Our description of the adsorption geometry of the pentacene molecule on Cu(511), Cu(911), and Cu(221) will focus on two characteristics: bending and tilting. Pentacene is planar in the gas phase. We find that in some cases, pentacene bends upon adsorption. We quantify this change in the molecular geometry by calculating the distance between the carbon atoms at the center and ends of the molecule, these values ($\Delta z$) can be found in Table 1. The tilting ($\alpha$) of the pentacene molecule reported in Table 1 was calculated with respect to the direction shown by the white arrow in Figure 2, which points toward the step edge. In Figure 2 we present a qualitative analysis of the changes in the molecular geometry of pentacene upon adsorption. The color assigned to each atom represents that atom's height above the copper surface and serves as an aide in our discussion of the adsorption geometries.

We found the greatest bending to occur for the adsorption of pentacene on Cu(511). On the other hand, pentacene was the least bent over the Cu(911) surface. The bending of pentacene on Cu(511) is strongly asymmetric, with one end of the molecule having a larger adsorption height than the other end. The optB88-vdW functional predicts less bending when compared to PBE for Pc/Cu(511), greater bending for Pc/Cu(221), and no change in the bending for the adsorption on Cu(911).

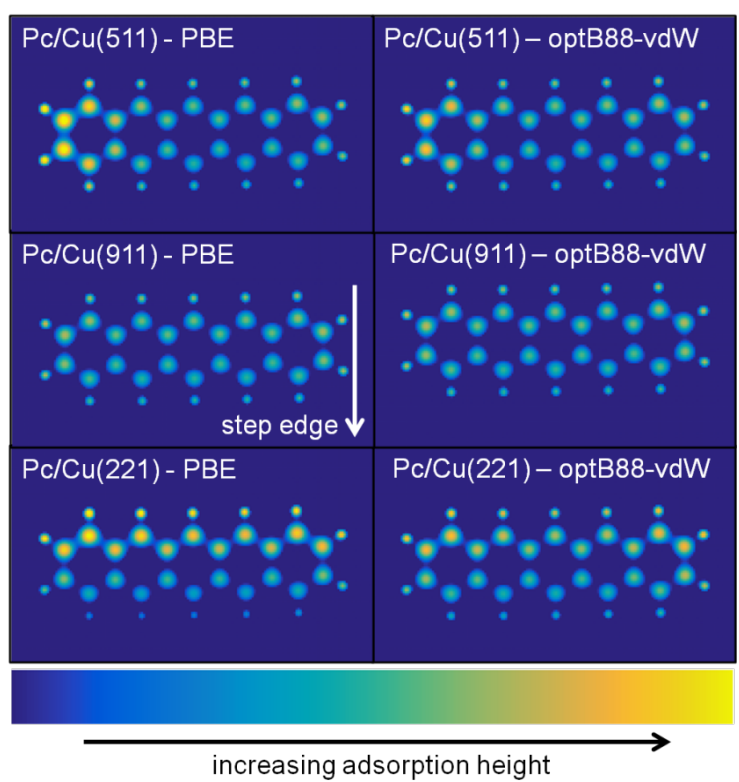

**Figure 2.** Representation of the differences in adsorption geometries of pentacene on Cu(511), Cu(911), and Cu(221) using the PBE and optB88-vdW functionals. The color map shows the relative adsorption heights of individual atoms in the pentacene molecule.

In all cases, pentacene is tilted such that the side that is closest to the step edge is closer to the surface (negative value with respect to the white arrow in Figure 2). The magnitude of this angle is the greatest for the adsorption of pentacene on Cu(221). With the addition of vdW interactions the tilt angle changes from -20.4° to -9.3°. For the adsorption of pentacene on Cu(511) and Cu(911), the angle is about -4° and decreases very slightly with the addition of vdW interactions.

In summary, the optB88-vdW functional predicts a flatter adsorption of pentacene on these vicinal surfaces of copper than the PBE functional. For the case of Pc/Cu(511), we find a reduction in the bending of pentacene when using optB88-vdW. On Cu(221) the bending of pentacene increases slightly with optB88-vdW, but the tilting is reduced by a large amount. The adsorption of pentacene on Cu(911) presents the smallest difference in adsorption geometry with the addition of vdW interactions, since pentacene is already at an almost flat adsorption geometry using the PBE functional.

Finally, we report the changes to the substrate first layer copper atoms upon the adsorption of pentacene (buckling). Large changes to the positions of first layer copper atoms are indicative of strong molecule-support interaction. We define buckling as the largest difference between the heights of any two surface copper atoms with respect to the clean surface plane. These values are reported in Table 1 for the adsorption on Cu(511), Cu(911), and Cu(221) using the PBE and optB88-vdW functionals. We find small, but non-negligible values of buckling for all of the cases considered here, ranging between 0.15 and 0.18 Å.

**III.2. Electronic Structure.**

Changes in the electronic structure of the substrate atoms, in the presence of adsorbates, are indicators of the molecule-substrate bonding. For this reason, the analysis of the electronic structure is an important part of a multifaceted approach aimed toward understanding the role of long-range interactions on the description of the adsorption as well as for differentiating between physisorption and chemisorption. Our analysis of the changes in electronic structure consists of an examination of changes in the d-band center and width, changes in the work function of the support, and charge transfer to the molecule in the for making these determinations.

**Table 2.** Summary of changes to the electronic structure upon adsorption of pentacene on Cu(511), Cu(911), and Cu(221) using the PBE and optB88-vdW functional. Shown from left to right are the change in the d-band center ($\Delta E_d$), change in the d-band full width at half maximum ($\Delta W_d$), change in the work function ($\Delta \Phi$), and charge transfer to the molecule from the support.

| Method | Surface | $\Delta E_d$ (eV) | $\Delta W_d$ (eV) | $\Delta \Phi$ (eV) | Charge Transfer (e-) |
|---|---|---|---|---|---|
| PBE | (511) | -0.28 | 0.57 | -0.57 | 1.0 |
|  | (911) | -0.40 | 0.53 | -0.64 | 1.0 |
|  | (221) | -0.23 | 0.50 | -0.35 | 0.8 |
| optB88-vdW | (511) | -0.28 | 0.55 | -0.64 | 0.9 |
|  | (911) | -0.37 | 0.55 | -0.61 | 0.9 |
|  | (221) | -0.25 | 0.45 | -0.41 | 0.8 |

The results of our electronic structure analysis are summarized in Table 2. Figures 3 and 4 show the d-band densities of state (d-DOS), for two rows of copper atoms under the pentacene molecule, calculated using PBE and optB88-vdW respectively. The change in the d-band center ($\Delta E_d$) and change in the full width at half maximum ($\Delta W_d$) are reported in Table 2 and extracted from the results plotted in figures 3 and 4. For all cases considered here, the adsorption of pentacene results in a shift in the d-band center away from the Fermi energy level and an increase in the d-band width as compared to the clean surface. The greatest shift in the d-band center corresponds to the adsorption on Cu(911), while the smallest shift corresponds to Pc/Cu(221).

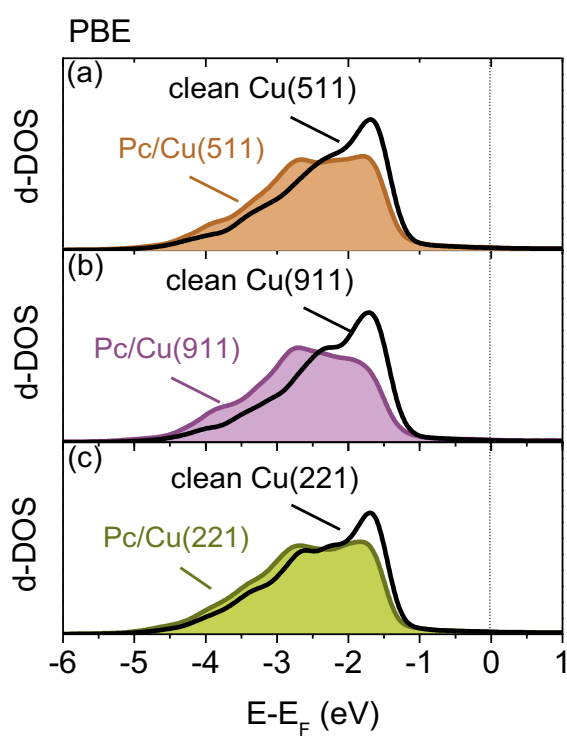

**Figure 3**. d-band densities of state (d-DOS) for the adsorption of pentacene on (a) Cu(511), (b) Cu(911) and (c) Cu(221) calculated using the PBE functional. The d-DOS for the clean support is shown in black.

The results for the change in the work function upon the adsorption of pentacene on Cu(511), Cu(911), and Cu(221) using PBE and optB88-vdW can also be found in Table 2. The calculated work function for the clean surfaces are in agreement with a previous investigation.[35] We calculate the change in the work function as the work function of the Pc/Cu system subtracted from the clean Cu surface. Therefore, negative changes in the work function represent a reduction of the work function with respect to the clean copper surface. We report a reduction in the work function of copper upon the adsorption of pentacene for all the surfaces and computational methods employed. We find the greatest change in the work function (-0.64 eV) for Pc/Cu(511) using optB88-vdW and Pc/Cu(911) using PBE. Additionally, the change in the work function is greater for pentacene on Cu(511) and Cu(911) when compared to that on Cu(221).

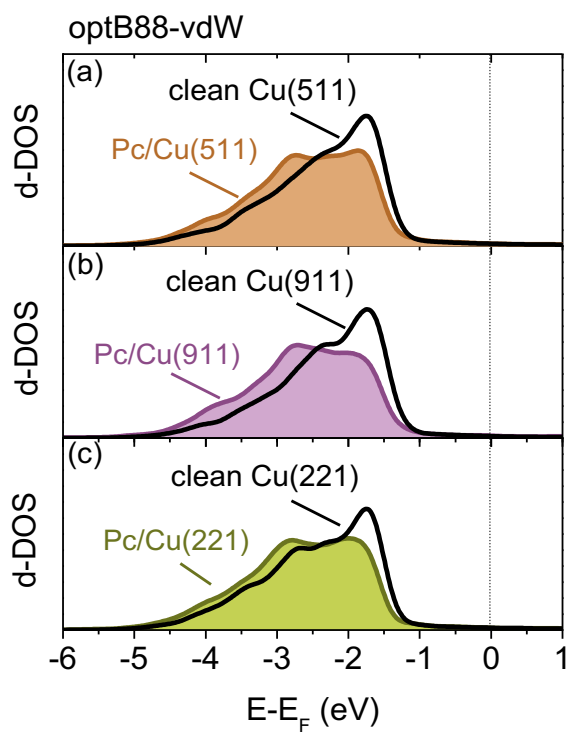

**Figure 4**. d-band densities of state (d-DOS) for the adsorption of pentacene on (a) Cu(511), (b) Cu(911) and (c) Cu(221) calculated using the optB88-vdW functional. The d-DOS for the clean support is shown in black.

We use the Bader method for calculating the charge transfer between the support and the pentacene molecule. In every case we find a transfer of electrons to the molecule from the support. In general we find between 0.8 and 1.0 electrons transferred to the pentacene molecule, constituting a significant net charge transfer (see Table 2). The PBE functional predicts a slightly higher charge transfer than the optB88-vdW functional for Pc/Cu(511) and Pc/Cu(911). The charge transfer found here is consistent with the charge transfer reported for the case of pentacene on Cu(110). [14]

## IV. Conclusion

We have studied the adsorption of pentacene on the vicinal surfaces of copper Cu(221), Cu(511) and Cu(911) using vdW inclusive DFT. We found a strong binding of pentacene to all of the supports that was enhanced by the inclusion of van der Waals interactions. Our calculations predict changes in the shape of the molecule upon adsorption, which include bending of pentacene on Cu(511) and a large tilting on Cu(221). The adsorption of pentacene caused changes to the electronic structure of the copper surface: a shift in the d-band center away from the Fermi level with respect to the clean surface, increase in the width of the d-band, and a net charge transfer to the molecule of about 1 electron. Aside from a strong enhancement in the adsorption energy, the addition of vdW did not cause a significant and systematic change in the adsorption characteristics of these systems, with the exception of the ~10˚ change in the tilt angle predicted by the optB88-vdW functional for pentacene on Cu(221). Based on these results and the criteria for chemisorption referenced earlier in the text, we suggest that pentacene on Cu(221), Cu(511) and Cu(911) are weakly chemisorbed systems.

## ACKNOWLEDGMENTS

This work was supported by the U.S. Department of Energy Basic Energy Science under Contract No DE-FG02-11ER16243. This research used resources of the National Energy Research Scientific Computing Center, which is supported by the Office of Science of the U.S. Department of Energy. We would like to thank Dr. Handan Yildirim for fruitful discussions.

**Corresponding Author:**

*E-mail: Abdelkader.Kara@ucf.edu.